\documentclass{iau}

\usepackage{amsmath}
\usepackage{graphicx}
\usepackage{multirow}

\begin{document}

\lefttitle{Solar photospheric spectrum microvariability}
\righttitle{Dainis Dravins \& Hans-G\"{u}nter Ludwig}

\jnlPage{1}{7}
\jnlDoiYr{2025}
\doival{10.1017/xxxxx}

\aopheadtitle{Proceedings IAU Symposium 400, Solar and Stellar Multi-Scale Activity,  }
\editors{tbd, eds.}

\title{Solar photospheric spectrum microvariability}

\author{{Dainis Dravins (1) and Hans-G\"{u}nter Ludwig (2)}}
\affiliation{(1) Lund Observatory, Division of Astrophysics, Department of Physics,\\ Lund University, SE-22100 Lund, Sweden\\ (2) Zentrum f\"{u}r Astronomie der Universit\"{a}t Heidelberg, Landessternwarte, \\ K\"{o}nigstuhl 12, DE--69117 Heidelberg, Germany\\}

\begin{abstract}
Finding low-mass planets around solar-type stars requires to understand the physical variability of the host star, which greatly exceeds the planet-induced radial-velocity modulation. This project aims at analyzing -- observationally and theoretically -- the character and physical origins of fluctuations in solar photospheric absorption lines.

Observationally, photospheric equivalent-width variations were measured in 1000 selected spectra from three years of HARPS-N data of the Sun-as-a-star, showing changes that largely shadow the chromospheric Ca~II~H\&K activity-cycle signal, but with much smaller amplitudes on sub-percent levels. Among iron lines, the greatest are for Fe~II in the blue, while the trends change sign among lines in the green Mg~I triplet and between Balmer lines.  No variation was seen in the semi-forbidden Mg~I 457.1 nm.

Theoretically, hydrodynamic 3D modeling of solar surface convection produced time sequences of synthetic high-resolution spectral atlases. Radial velocities averaged over small simulation areas jitter by some {$\pm$}150~m\,s$^{-1}$, scaling to $\sim${$\pm$}2~m\,s$^{-1}$ for the full solar disk on timescales of granular convection.  Among different lines, jittering is in phase, but amplitudes differ by about one tenth of their values: greater for stronger and for ionized lines, decreasing at longer wavelengths 
\end{abstract}

\begin{keywords}
Sun: spectrum, Sun: photosphere, Exoplanets: radial velocities, Exoplanets: terrestrial planets
\end{keywords}

\maketitle

\section{Introduction}

Efforts are underway toward enabling the detection of ´exoEarths', i.e., planets of about one Earth mass, in about one-year orbits around solar-type stars.  The most promising method appears to be the radial-velocity one, with instrumental precisions now approaching 10~cm\,s$^{-1}$ \citep{crassetal21, fischeretal16, rackhametal23}.  However, such an exoplanet signal is overwhelmed by intrinsic stellar variability, where the currently best modeling is unable to extract planetary signals with amplitudes much below 1 or 2 m\,s$^{-1}$.  To reach adequate sensitivity thus demands to understand the complexities of stellar atmospheric dynamics and spectral line formation, manifest as a physical jitter or drift of its apparent radial velocity.  A step toward exoEarth detection could be to identify dissimilar spectral lines (e.g., strong or weak, neutral or ionized, high or low excitation, atomic or molecular, short or long wavelength, magnetically sensitive or not), with disparate responses to stellar activity, which are modulated in some characteristic manner, to disentangle wavelength shifts induced by exoplanets from those originating in solar-type atmospheres.

As measures of stellar activity, chromospheric indices are often used, in particular Ca~II~H\&K.  However, precision radial velocities are measured in photospheric lines only, for which chromospheric indices at best can be only an approximation.  For example, while fluctuations in apparent solar photospheric radial velocities overall do correlate with Ca~II~H\&K indices, there is a time lag of several days between them, seemingly due to the photospheric signal being more dependent on the areal extent of magnetic plage regions than on centers of chromospheric emission \citep{bruninglabonte85, colliercameronetal19}, requiring studies of microvariability in specifically photospheric absorption lines. 

This project studies the properties and origins of fluctuations in different classes of absorption lines in solar-type spectra.  To develop optimal algorithms for the extraction of radial velocities from observational data, will most likely require a detailed understanding of how stellar physical variability affects different photospheric lines.  Besides leading to improved data extraction from current instruments, such an understanding should also provide guidance toward types of future instrumentation that may be required for exoEarth surveys.

\section{Line-strength variability during the solar cycle}

Past efforts to identify solar-cycle variability of photospheric lines in the spectrum of the Sun-as-a-star included 35 years of monitoring from Kitt Peak \citep{livingstonetal07, livingstonetal10}, but with rather inconclusive results except for chromospherically influenced lines, acknowledging the challenges in maintaining instrumentation and observational conditions stable over longer epochs with spectrometers operated in air and readjusted between different observing programs.  A review of these and other earlier efforts is in \citet{dravinsludwig24}. 

These limitations are now circumvented by Sun-as-a-star telescopes feeding extreme-precision radial-velocity spectrometers operated in vacuum and (almost) without any readjustments.  Their time series permit not only radial-velocity monitoring but also precise stellar spectroscopy.

\subsection{HARPS-N observations of the Sun-as-a-star}

On La Palma, a small sunlight-integrating telescope feeding the HARPS-N spectrometer continues to keep observing the Sun-as-a-star.  Its first public data release covered the three-year period from July 2015 through July 2018, a declining phase of the solar activity cycle \citep{colliercameronetal19, dumusqueetal21, milbourneetal19}.  Those spectra, with a spectral resolution $\lambda$/$\Delta\lambda$\,$\sim$115,000, cover the wavelength interval of 387-691 nm.

Out of some 35,000 recordings in that dataset, 1000 spectra were selected for best photometric signal-to-noise (typically $\sim$350) and smallest airmass (minimizing effects by telluric absorption), singling out summertime observations around daily noon (for the La Palma latitude, then close to zenith) during 2016, 2017 and 2018.  Groups of largely unblended lines were chosen: Fe~I of varying strengths between 430--685 nm; Fe~II lines in the blue, stronger lines such as the green Mg~I triplet, the semi-forbidden intercombination line Mg~I 457.1 nm, the molecular G-band, hyperfine split Mn~I lines, and also the chromospherically influenced Balmer lines, Na~I D$_1$ and D$_2$. For details, see \citet{dravinsludwig24}.

\subsection{Photospheric line changes during successive years}

Figure 1 shows systematic patterns in measured line-strength variations.  With increasing chromospheric activity, Fe absorption gradually weakens, with the greatest amplitude for Fe~II at shorter wavelengths, diminishing for Fe~I, and for longer wavelengths. 

For stronger, chromospherically influenced lines such as the green triplet of Mg~I b$_1$ 518.3, b$_2$ 517.2, and b$_3$ 516.7 nm, the behavior is more complex, and the changes are systematically different among lines in this triplet.  Their atomic energy levels are coupled by being transitions between one common upper level and three different lower levels, a configuration that makes non-LTE effects very likely. Somewhat similar relations are seen among the H$\alpha$, H$\beta$ and H$\gamma$ Balmer lines of hydrogen.  By contrast, the formation of the semi-forbidden Mg~I 457.1 nm intercombination line is expected to be relatively simple, being dominated by atomic collisions.  Indeed, its behavior is very different, showing no significant variability during these years.

  \begin{figure}
  \centering
    \includegraphics[scale=.51]{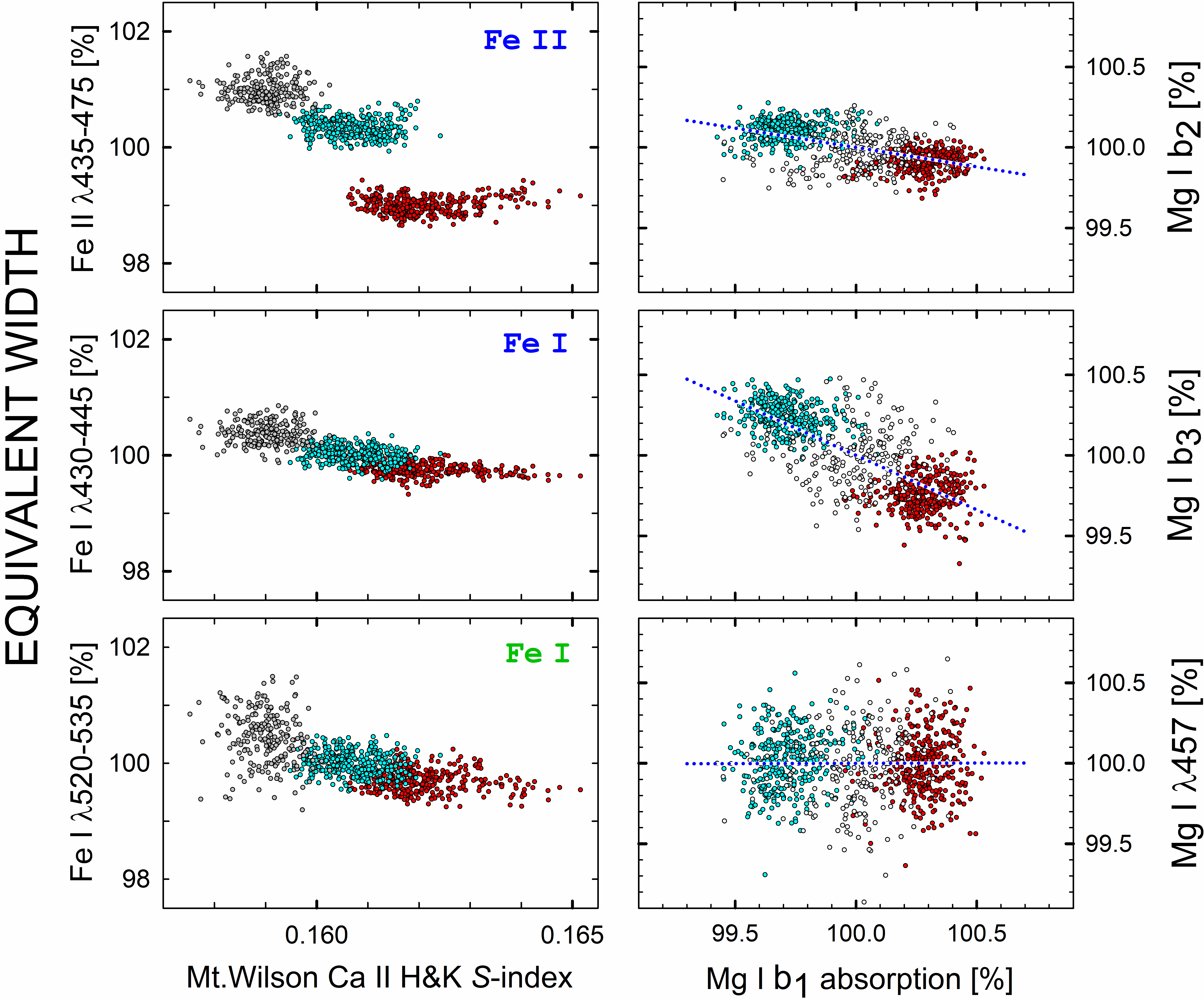}
    \caption{Left: Observed relative changes of Fe~I and Fe~II absorption line equivalent widths in the spectrum of the Sun-as-a-star, measured with HARPS-N, between summer seasons in 2016 (red), 2017 (cyan), and 2018 (gray), as a function of the Mt.Wilson Ca~II~H\&K chromospheric activity S-index.  The variability largely shadows the Ca~II~H\&K activity-cycle signal, but with much smaller amplitudes on sub-percent levels. Right: Relative and systematic changes in the absorption equivalent widths between the three strong lines of the Mg~I green triplet, and the semiforbidden Mg~I $\lambda$457 nm line. }
    \label{image one}
  \end{figure}

Such patterns of photospheric line variability demonstrate that these contain nontrivial information and measure something different than chromospheric indices.  However, detailed studies of such signals on their sub-percent levels remain challenging.  With equivalent-width measures formed over many tens of pixels in the spectrometer, each with nominal S/N $\sim$350, and after averaging over several similar lines, one would expect the photometric noise contribution to be less than 0.001, raising the question what part of the scatter seen in Fig.\,1 is physical and what is instrumental?  

It would certainly be interesting to follow the relative radial-velocity and line-shape changes between groups of different classes of lines during the solar activity cycle; however that remains beyond the capabilities of current radial-velocity instruments.  Their wavelength calibration is stable but not absolute between different grating orders.  They reach sub-m\,s$^{-1}$ precision, but only after averaging over complete stellar spectra, not for limited groups of selected lines.  Although their spectral resolutions are classed as `high', those are lower than in synthetic spectra or in spectrometers at solar telescopes, smearing out signatures such as the exact bisector shapes. For HARPS-N and similar instruments, such limitations are discussed by \citet{hanassisavarietal25}.  For such reasons, the current studies concern equivalent widths only, which -- at least to a first approximation -- should be insensitive to such issues. 

Several precision radial-velocity instruments are used to monitor the Sun-as-as-star and comparisons among them confirm that their deduced radial-velocity fluctuations largely agree \citet{zhaoetal23}.  With time series now reaching a full solar cycle, it should be feasible to examine whether also deduced line-strength variations agree among the instruments, then isolating what is solar and what is instrumental.

\section{Theoretical modeling of radial-velocity jittering}

Quiet  granulation covers most of the solar surface and provides most of the solar flux.  For its modeling, surface convection is simulated with time-dependent 3D hydrodynamics, along with the concurrent computation of complete spectra from ultraviolet to the infrared.  With a hyper-high spectral resolution, the changing shapes and wavelength shifts of representative Fe~I and Fe~II lines are followed during the simulation sequences, seeking systematic patterns of variability, in particular such that correlate with excursions in apparent radial velocity.

Solar surface volumes that realistically can be simulated cover only a small surface fraction: in the present modeling with the CO\,$^5$BOLD code this is a 3D Cartesian “local-box” model of 5.6\,$\times$\,5.6 Mm$^2$; for modeling methodology, see \citet{freytagetal12}.  From this simulation volume, spatially averaged line profiles are computed in LTE at various emergent angles, and corresponding full-disk profiles synthesized for a non-rotating Sun.  The radial-velocity jittering for integrated sunlight, built up from such a small simulation area, typically amounts to $\pm$\,150~m\,s$^{-1}$, scaling to $\sim$2~m\,s$^{-1}$ for the real full solar disk.  Most photospheric lines vary in phase but differences in radial-velocity excursions between various line-groups are found to be about 10\% of their values.  Amplitudes are greater for stronger and for ionized lines, decreasing at longer wavelengths.  An example is in Fig.\,2 with details in \citet{dravinsludwig23}.

Current computational constraints do not permit to properly make detailed simulations of granulation coupled to large-scale features such as supergranular and meridional flows, p-mode oscillations, and other, which certainly may contribute significant radial-velocity signals but would require much larger simulation volumes. 

 \begin{figure}
   \centering
    \includegraphics[scale=.45]{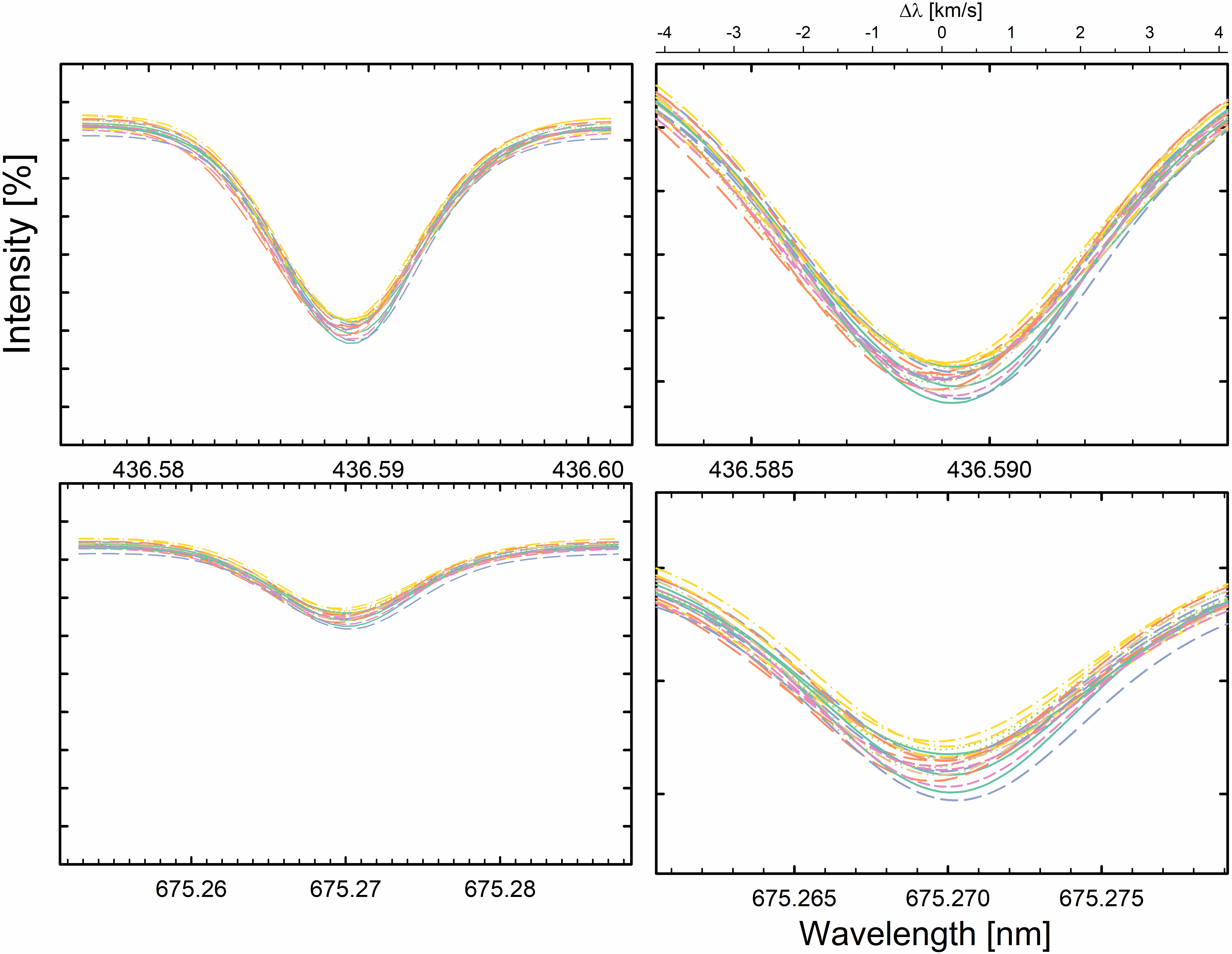}
    \caption{Synthetic spectral lines for the spectrum of a full solar disk, built up by spectral-line components from a small simulation area of non-magnetic granulation, sampled at 20 instances during the 3D hydrodynamic simulation.  Top: A strong Fe~I line in the blue, $\lambda$436.59 nm, $\chi$\,=\,2.99 eV. Bottom: A weak Fe~I line in the longward red, $\lambda$675.27 nm, $\chi$\,=\,4.64 eV.  Right: zoomed-up line centers on a uniform wavelength scale in units of radial velocity (top); tick marks on the vertical axes correspond to those for the full profiles. Hyper-high spectral resolution, $\lambda$/$\Delta\lambda$ $\sim$900,000. }
    \label{image two}
  \end{figure}

\subsection{Possible mechanism of differences between line-groups}

For solar-type stars, the greatest convective blueshifts are found in weak lines formed in the deeper photosphere.  However, the variability is found to be smallest in these weak lines,  increasing for stronger lines formed in the upper photosphere.  A typical 3D atmospheric structure, e.g., Fig.\,14  in \citet{nordlundetal09} shows the smallest temperature fluctuations to occur at optical depths around $\tau$\,=\,0.3, reaching greater values both deeper down and higher up.  Weaker spectral lines form around about $\tau$\,=\,0.5, where the fluctuations are smaller than on the more elevated formation heights of the stronger lines with lower atmospheric densities, where the gases may roam more freely and lines may experience greater jittering amplitudes. 

\subsection{Testing model predictions?} 

Due to their smallness, some model predictions are not easily verified in current observations of the Sun-as-a-star.  However, the modeling may be tested with spatially resolved observations across the solar disk.  The radial-velocity jittering is predicted to be greatest at disk center, gradually decreasing toward the limb, with different behavior for stronger and weaker lines, and for those of different excitation potentials \citep{dravinsludwig23}. While the disk-integrated signal amplitudes differ by only some fraction of a m\,s$^{-1}$, the signal over an area of a few arcminutes is greater by orders of magnitude, and should be realistic to sample.

\section{Magnetic granulation}

The second largest contribution to solar irradiance comes from plage and faculae areas of magnetic granulation. Magnetic fields disturb and dampen the convective velocity patterns, causing different asymmetries of the line profiles and resulting in much smaller convective blueshifts. The varying area coverage of magnetic granulation during the solar cycle has been identified as the main driver for long-term changes in solar apparent radial velocity \citep{meunieretal10a, meunieretal10b}.  While contributions from sunspots also affect spectral lines \citep{komorietal25}, their fraction of the solar photospheric flux is modest.

While non-magnetic granulation in full-disk spectra, in principle, can be modeled from fundamental physical parameters of stellar temperature, gravity and chemical composition, there are more degrees of freedom for magnetic granulation.  Not only can their area coverage and latitudinal and longitudinal surface distributions be different (and unknown for other stars), but also the degree and distribution of magnetic flux concentrations.  For the spectral synthesis, an additional complication (or opportunity!) comes with the potential of handling the components of magnetically sensitive lines, in a situation with a sometimes uneven precision of laboratory data.
  
Nevertheless, line-profile variability in magnetic areas could reveal signatures that might discriminate against other effects.  A sequence of magnetohydrodynamic CO\,$^5$BOLD solar models were computed as above, representing a range of magnetic flux densities. The vertical magnetic flux through the atmosphere was fixed by initial conditions and preserved during the modeling sequence.  Fieldlines were constrained to pass vertically through the top and bottom boundaries of the simulation volume.  For details, including synthetic surface images for different magnetic flux levels and at different limb angles, see \citet{ludwigetal23}.  Synthetic LTE spectra for different flux densities were computed at numerous instances throughout the simulations, and Fig.\,3 provides some examples for 2400 G (240 mT), compared to the non-magnetic case.  

 \begin{figure}
   \centering
    \includegraphics[scale=.45]{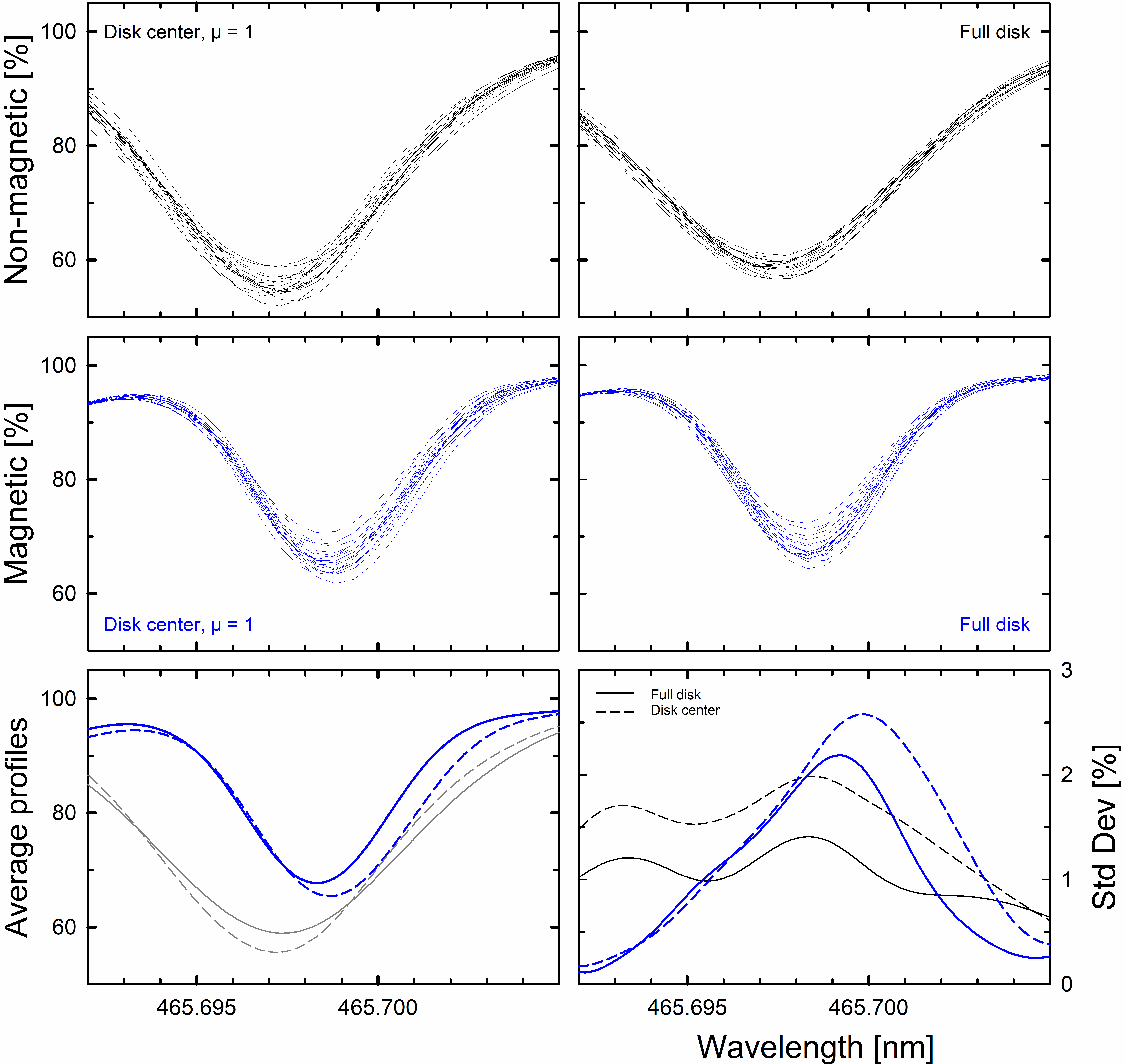}
    \caption{Synthetic solar profiles for an ordinary Fe~II line ($\lambda$\,=\,465.698 nm, $\chi$\,=\,2.89 eV) in non-magnetic and magnetic (2400 G, 240 mT) granulation.  Thin curves are spatially averaged profiles across the simulation area, sampled at 20 temporal instances; black for non-magnetic; blue for magnetic.  Left-hand groups show profiles at solar disk center, at right integrated for a full (non-rotating) solar disk, using center-to-limb profiles from the small simulation area.  Spatially and temporally averaged profiles are at bottom left.  The non-magnetic convective blueshift is greater (by $\sim$\,400~m\,s$^{-1}$), and furthermore is slightly larger at disk center than in integrated sunlight.  The opposite holds for magnetic granulation, apparently caused by a more `corrugated' stellar surface, an effect also causing increased plage and faculae brightness toward the limb \citep{ludwigetal23}. Bottom right: Hyper-high spectral resolution ($\lambda$/$\Delta\lambda$ $\sim$900,000) enables to study variations inside line profiles. The temporal standard deviations show the longward line flanks to be the most variable in magnetic granulation versus the line core in the non-magnetic case. }
    \label{image three}
  \end{figure}

\section{Outlook: possibilities and challenges}

A significant addition to Sun-as-as-star facilities will be the PoET telescope, to be installed at ESO VLT on Paranal \citep{santosetal25}.  Besides providing spectra of the Sun-as-a-star with the ESPRESSO spectrometer's higher resolving power of $\sim$200,000, it will be able to select limited portions of the solar disk to examine local radial velocities.  This should help to identify locations contributing to solar velocity jittering and also enable novel tests of hydrodynamic model atmospheres.  Until now, these have been verified against spatially and temporally averaged spectral line shapes, granulation morphology, and intensity modulation.  Predictions such as different radial velocity jittering amplitudes among various classes of spectral lines -- such as discussed above -- cannot yet be tested because of lack of adequate instrumentation.  For Sun-as-a-star, it is still not possible to discern differences between line-group jittering of amplitudes between, say, 2.0 and 2.2~m\,s$^{-1}$, as deduced for the full solar disk.  However, such extrapolated full-disk values emanate from small simulation areas, where the amplitudes are much greater and should be readily measurable with ESPRESSO. 

Different levels of approximation and sophistication are possible in computing synthetic spectral lines from dynamic 3-D atmospheres, eventually perhaps envisioning `full' non-LTE treatments with Zeeman-sensitive line components, a connection to the lower chromosphere, and other.  Although all such are not yet feasible, the types of observations foreseen with PoET may well guide toward highly precise 3-D spectra.  To realize corresponding spatially resolved spectroscopy across surfaces of other stars is marginally possible using exoplanet transits to differentially isolate surface portions, for which theoretical line-profile predictions can be made \citep{dravinsetal17a, dravinsetal17b, dravinsetal18, dravinsetal21a, dravinsetal21b}.  In the nearish future, the availability of stable spectrometers at also extremely large telescopes should make such observations less challenging. 

The significance of non-LTE effects for also `ordinary' photospheric lines measured for radial velocities is suggested by studies of how different assumed levels of chromospherically influenced temperatures in the upper photosphere affect 1-D non-LTE line formation in modeled G2~V stars \citep{vieytesetal25}.  The response to different levels of chromospheric heating is markedly different among various Fe~I lines, even such with similar strengths at nearby wavelengths, carrying a promise that those may contain additional and perhaps unique signatures of solar atmospheric modulation.  Such lines could be selected to observationally search for their variability although challenging non-LTE 3D calculations might be needed for their fuller understanding.  

Other relevant spectrometers for Sun-as-a-star spectroscopy include those in the near infrared, e.g.,  NIRPS at ESO on La Silla \citep{bouchyetal25}.  As seen in both observed and synthetic spectra, the amplitudes of physical variability decrease toward longer wavelengths, suggesting that the near infrared could be advantageous for exoEarth searches.  However, there is the complexity of fewer atomic lines of significant strength in that region, and effects from enhanced telluric absorption.

The large ANDES spectrometer is being built for the ESO ELT \citep{palleetal25}.  While it of course will have outstanding light-collecting power, and should provide very high S/N spectra for brighter stars, the challenges of interfacing a realistically large cross-dispersed échelle spectrometer to an extremely large telescope, limits its spectral resolution to similar ranges as current instruments.  In the longer term, one could certainly wish for also hyper-high resolutions of 1,000,000, say, then possibly requiring diffraction-limited adaptive-optics photonic instruments to limit their physical size. Already NIRPS is taking advantage of adaptive optics to limit its size but to realize the same for an extremely large telescope in the shorter visual wavelengths remains an interesting challenge.

\begin{acknowledgements}
{The work by DD is supported by grants from The Royal Physiographic Society of Lund.  Extensive use was made of NASA’s ADS Bibliographic Services and the arXiv distribution service. }

\end{acknowledgements}

\end{document}